\documentclass{article}
\newcommand{\be}{\begin{equation}}
\newcommand{\ee}{\end{equation}}
\newcommand{\bea}{\begin{eqnarray}}
\newcommand{\eea}{\end{eqnarray}}
\newcommand{\al}{\alpha}
\begin{document}
\title{Understanding Heisenberg's `magical' paper of July 1925: a new look at  the calculational details}
\author{Ian J. R. Aitchison \\
 {\normalsize  Department of Physics, Theoretical Physics, University of Oxford,} \\
{\normalsize Oxford OX1 3NP, UK}\\ \\
David A. MacManus\\
{\normalsize Tripos Receptor Research Ltd., Bude-Stratton Business Park,}\\
{\normalsize Bude, Cornwall EX23 8LY, UK}\\ \\
Thomas M. Snyder\\ 
{\normalsize Department of Mathematics and Engineering Sciences, Lincoln Land Community College,} \\
{\normalsize Springfield, Illinois 62794-9256}}
\maketitle

\begin{abstract}
In July 1925 Heisenberg published a paper [Z. Phys. {\bf 33} 879-893 (1925)] 
which ended the period of `the Old Quantum Theory' and ushered in the new era 
of Quantum Mechanics. This epoch-making paper is generally  regarded 
as being difficult to follow, perhaps  
 partly because Heisenberg provided few clues as to how he arrived 
at the results which he reported. Here we give details of calculations of the 
type which, we suggest, Heisenberg may have performed. 
We  take as a specific example one of the anharmonic oscillator problems considered by Heisenberg, and use  
our reconstruction of his  approach
 to solve it up to second order in perturbation theory. 
We emphasize that the results are precisely those obtained in standard 
quantum mechanics, and suggest that some discussion of the approach - based 
on the direct computation of transition amplitudes - could usefully be 
included in undergraduate courses on quantum mechanics. 
 
\end{abstract}

\section{Introduction}

Heisenberg's paper of July 1925$^1$, entitled `Quantum-mechanical re-interpretation of kinematic and mechanical relations'$^{2,3}$, was 
the breakthrough which quickly led to the first complete formulations of 
quantum mechanics$^{4,5,6}$. Despite its undoubtedly crucial historical role, 
Heisenberg's paper is not generally referred to in undergraduate courses on 
quantum mechanics - in contrast, say, to the place of Einstein's 1905 paper 
in the teaching of Relativity. Indeed it is still widely regarded as 
being difficult to understand and  - perhaps because of this -  of  only 
historical interest today. For example, Weinberg$^7$ has written as follows: 

`If the 
reader is mystified at what Heisenberg was doing, he or she is not alone. I have tried 
several times to read the paper that Heisenberg wrote on returning from 
Heligoland, and, although I think I understand quantum mechanics, I have never 
understood Heisenberg's motivations for the mathematical steps in his paper. Theoretical 
physicists in their most successful work tend to play one of two roles: they are 
either {\em sages} or {\em magicians}....It is usually not difficult to understand the 
papers of sage-physicists, but the papers of magician-physicists are often 
incomprehensible. In that sense, Heisenberg's 1925 paper was pure magic. 

Perhaps we should not look too closely at Heisenberg's first paper......'

There have, in fact, been many discussions aimed at elucidating the main 
ideas in Heisenberg's paper, of which Refs 3, 8, 9-18 no doubt represent only 
a partial selection; of these the most detailed appear to be Ref. 3 pages 28-35, 
 Ref. 8 pages 204-224, Ref. 10 pages 161-188, and Ref. 11 chapter IV.  Of course, it may well not be possible ever to render 
completely comprehensible the mysterious processes whereby magician-physicists 
`jump over all intermediate steps to a new insight about nature'$^{19}$. In our 
opinion, however, one of the main barriers to understanding Heisenberg's paper, for 
most people, is a more prosaic one: namely, that he gives remarkably few details 
of the calculations he actually performed, in order to arrive at his results  
for the one-dimensional model systems which he treats (anharmonic oscillators 
and the rigid rotator). 

One aim of the present paper is therefore to make Heisenberg's paper more 
accessible to scientists, and to historians of science, by briefly reviewing 
the line of reasoning he followed in setting up his new calculational scheme 
(section 2), and then by presenting (in section 3) full details of a 
calculation typical of those we conjecture that Heisenberg himself 
performed. Our `reconstruction' is based on the assumption that,  having 
formulated a scheme which was capable in principle of determining uniquely the 
relevant physical quantities (transition frequencies and amplitudes), 
Heisenberg then applied it rigorously to various `toy' mechanical systems, 
without any further recourse to the kind of `inspired guesswork' that characterised 
the Old Quantum Theory. Surprisingly, this point of view appears to be novel: 
MacKinnon$^{10}$, and  
Mehra and Rechenberg$^{11}$, for example, consider that Heisenberg arrived at 
the crucial recursion relations (equations (\ref{eq:2.40}) - (\ref{eq:2.43}) below),  in the quantum case, 
by  essentially guessing the 
appropriate generalisation of their classical counterparts (see section 3). We 
are unaware of any evidence, now, that can settle the issue. In any case, our 
analysis shows that it is possible to read  Heisenberg's paper as 
providing  a complete (if limited)  calculational scheme,  
the results of applying which are precisely those of 
standard quantum mechanics. Thus a second aim of our paper is to stress both the 
correctness and the practicality of what we conjecture to be 
Heisenberg's calculational scheme, and to 
stimulate a re-appraisal of the possibility of including at least some discussion of 
it in undergraduate courses. 

\section{Heisenberg's `transition amplitude' approach} 

\subsection{Quantum kinematics}

As is well known, Heisenberg begins his paper with a programmatic call to 
`discard all hope of observing hitherto unobservable quantities, such as the 
position and period of the electron'$^{20}$, and instead to `try to establish 
a theoretical quantum mechanics, analogous to classical mechanics, but in 
which only relations between observable quantities occur'. As an example of such 
latter quantities, he immediately points to the energies $W(n)$ of the (Bohr) 
stationary states, together with the associated (Einstein-Bohr) frequencies$^{21}$
\be
\omega(n, n-\alpha) = \frac{1}{\hbar} \{W(n) - W(n-\alpha)\}, \label{eq:2.1}
\ee
noting that these frequencies, which characterise radiation emitted in the 
transition $n \to n-\alpha$, depend on two variables. An example of something 
he wishes to exclude from the new theory is the time-dependent position coordinate $x(t)$. In considering what might replace it, he turns to the 
 probabilities for transitions between stationary states. Consider a simple 
one-dimensional model of an atom consisting of an electron undergoing 
periodic motion, which is in fact the type of system studied by Heisenberg. 
For a state characterised by the label $n$, fundamental frequency $\omega(n)$ 
and coordinate $x(n, t)$, one can represent $x(n, t)$ as a Fourier series 
\be 
x(n, t) = \sum_{\al = - \infty}^{\infty} X_{\al}(n) {\rm exp}[{\rm i}\omega(n) \al t].
\label{eq:2.4}
\ee 
According 
to classical theory, the energy emitted per unit time (that is, the power) 
in a transition  corresponding to the $\al$th harmonic $\omega(n)\al$ is$^{22}$ 
\be 
-\left( \frac{{\rm d} E}{{\rm d}t}\right)_{\alpha} = \frac{e^2}{3 \pi \epsilon_0 c^3} [\omega(n)\al]^4 |X_{\al}(n)|^2 . \label{eq:2.2}
\ee 
In the quantum theory, however, the transition frequency corresponding to the 
classical `$\omega(n)\al$' is in general {\em not} a simple multiple of a 
fundamental frequency, but is given by ({\ref{eq:2.1}): thus $\omega(n)\al$ 
is replaced by $\omega(n, n-\al)$. Correspondingly, Heisenberg introduces 
the quantum analogue of $X_\al(n)$, which he writes as $X(n, n-\al)^{24}$. 
Further, the left hand side of (\ref{eq:2.2}) has, in the quantum theory, to 
be replaced by the product of the transition probability per unit time, 
$P(n, n-\al)$, and the emitted energy $\hbar \omega(n, n-\al)$; thus (\ref{eq:2.2}) becomes 
\be
P(n, n-\al) = \frac{e^2}{3 \pi \epsilon_0 \hbar c^3} [\omega(n, n-\al)]^3 |X(n, n-\al)|^2. \label{eq:2.3}
\ee
It is the 
{\em transition amplitudes} $X(n, n - \al)$ which Heisenberg fastens upon 
as being satisfactorily `observable'; like the transition frequencies, they 
depend on two discrete variables$^{25}$. 

Equation (\ref{eq:2.3}) refers, however, to only one specific transition. For a full 
description of atomic dynamics (as then conceived), one will need to consider 
all the  quantities $X(n, n-\al){\rm exp}[{\rm i} \omega(n, n-\al)t]$. In the classical case, the  terms  $X_\al(n) {\rm exp}[{\rm i} \omega(n)\al t]$  may be combined to yield $x(t)$ via (\ref{eq:2.4}). But in 
the quantum theory, Heisenberg says, a `similar combination of the corresponding quantum-theoretical 
quantities seems to be impossible in a unique manner and therefore not meaningful, in view of the equal weight of the variables $n$ and $\al$ 
[i.e. in the amplitude $X(n, n-\al)$ and frequency $\omega(n, n-\al)$].' `However', he continues, `one may readily regard the ensemble of quantities 
$X(n, n-\al){\rm exp}[{\rm i} \omega(n, n-\al)t]$ as a representation of 
the quantity $x(t) \ldots$' This is the first of Heisenberg's `magical 
jumps' - and certainly a very large one. Representing $x(t)$ in this way seems 
to be the sense in which he considered himself to be `re-interpreting the 
kinematics'.

Still concerned with the kinematics, Heisenberg immediately poses the question: 
`how is the quantity $x(t)^2$ to be represented?' In classical theory, the 
answer is straightforward. From (\ref{eq:2.4}) we obtain 
\be 
[x(t)]^2 = \sum_\al \sum_\gamma X_\al(n) X_\gamma(n) {\rm e}^{{\rm i} \omega(n) (\al + \gamma) t}.
\label{eq:2.5}
\ee
Relabelling $\al + \gamma$ as $\beta$, (\ref{eq:2.5}) becomes 
\be 
[x(t)]^2 = \sum_\beta Y_\beta(n) {\rm e}^{{\rm i} \omega(n) \beta t}
\label{eq:2.6}
\ee
where
\be 
Y_\beta(n) = \sum_\al X_\al(n) X_{\beta - \al}(n).
\label{eq:2.7}
\ee
Thus classically $[x(t)]^2$ is represented (via a Fourier series) by the set of 
quantities $Y_\beta(n){\rm exp}[{\rm i} \omega(n) \beta t]$, the frequency 
$\omega(n) \beta$ being the simple combination $[\omega(n)\al + \omega(n)(\beta - \al)]$. In quantum theory, the corresponding representative quantities 
must be written as $Y(n, n-\beta){\rm exp}[{\rm i} \omega(n, n-\beta)t]$, and 
the question is: what is the analogue of (\ref{eq:2.7})?

The crucial difference in the quantum case is that the frequencies do {\em not} 
combine in the same way as the classical harmonics, but rather in accordance with the Ritz combination principle:
\be 
\omega(n, n-\al) + \omega(n-\al, n-\beta) = \omega(n, n-\beta), 
\label{eq:2.8}
\ee
which is of course consistent with (\ref{eq:2.1}). Thus in order to end up with 
the particular frequency $\omega(n, n-\beta)$, it seems `almost necessary' 
(in Heisenberg's words) to combine the quantum amplitudes in such a way as 
to ensure the frequency combination (\ref{eq:2.8}); that is, as 
\be 
Y(n, n-\beta) {\rm e}^{{\rm i} \omega(n, n-\beta)t} = \sum_{\al}X(n, n-\al) 
{\rm e}^{{\rm i} \omega(n, n-\al) t} X(n-\al, n-\beta) {\rm e}^{{\rm i} \omega(n-\al, n-\beta)t}.
\label{eq:2.9}
\ee
Cancelling the exponentials on both sides of (\ref{eq:2.9}) we are left with 
\be 
Y(n, n-\beta) = \sum_\al X(n, n-\al) X(n-\al, n- \beta), 
\label{eq:2.10}
\ee
which is Heisenberg's law for multiplying transition amplitudes together. 

He indicates the simple extension of the rule to higher powers $[x(t)]^n$, but 
at once notices that a `significant difficulty arises, however, if we consider 
two quantities $x(t), y(t)$ and ask after their product $x(t)y(t).\ldots$ 
Whereas in classical theory $x(t)y(t)$ is always equal to $y(t)x(t)$, this is 
not necessarily the case in quantum theory'. This `difficulty' clearly 
unsettled Heisenberg: but it very quickly became clear that the {\em non-commutativity} (in general) of kinematical quantities in quantum theory 
was the really essential new technical idea in the paper. Born recognised 
(\ref{eq:2.10}) as matrix multiplication (something unknown to 
Heisenberg in July 1925), and he and Jordan rapidly produced Ref. 4, the first paper 
to state the fundamental commutation relation (in modern notation) 
\be 
\hat{x} \hat{p} - \hat{p} \hat{x} = {\rm i} \hbar. 
\label{eq:2.11}
\ee
Dirac's paper followed soon after$^5$, and then the `three-man' paper of 
Born, Heisenberg and Jordan$^6$. 

The economy and force of Heisenberg's argument in reaching (\ref{eq:2.10}) 
is surely very remarkable - and it seems at least worth considering whether 
presenting it to undergraduates might not help them to understand the 
`almost necessity' of non-commuting quantities in quantum theory. 

\subsection{Quantum dynamics}

Having identified the transition amplitudes $X(n, n-\al)$ and frequencies 
$\omega(n, n-\al)$ as the `observables' with which the new theory should 
deal, Heisenberg now turns his attention to how they may be determined 
`from the given forces of the system' - that is, by the dynamics. In the 
Old Quantum Theory, he notes, this would be done in two stages: (a) by 
integration of the equation of motion 
\be 
\ddot{x} + f(x) = 0 \label{eq:2.12},
\ee  
 and (b) by determining the constants of periodic motion through the 
`quantum condition'
\be
\oint p {\rm d} q = \oint m {\dot{x}}^2 {\rm d} t \equiv J (=n h),
\label{eq:2.13}
\ee
where the integral is to be evaluated over one period of the motion. 
As regards (\ref{eq:2.12}), Heisenberg says that it is `very natural' to take 
over the classical equation of motion into quantum theory, replacing the 
classical quantities $x(t)$ and $f(x)$ by their kinematical re-interpretations$^{26}$, as in the previous section (or, as we would say 
today, by taking matrix elements of the operator equation of motion 
$\ddot{\hat{x}} + f(\hat{x}) = 0$). He notes that in the classical case a 
solution can be obtained by expressing $x(t)$ as a Fourier series, insertion 
of which into the equation of motion leads (in simple special cases) 
to a set of recursion relations for the Fourier coefficients. 
In the quantum theory, 
Heisenberg says, `we are at present forced to adopt this way of solving 
equation (\ref{eq:2.12}) [his equation H(11)]....since it was not possible 
to define a quantum-theoretical function analogous to the [classical] function 
$x(n, t)$.' In section 3 we shall consider the simple 
example (the first of those chosen by Heisenberg) $f(x)=\omega^2_0 x + \lambda x^2$, obtaining the appropriate recursion relations in both the 
classical and the quantum cases.

A `quantum-theoretical re-interpretation' of (\ref{eq:2.13}) is now 
required, in terms of the transition amplitudes $X(n, n-\al)$. In the classical case, insertion of (\ref{eq:2.4}) into (\ref{eq:2.13}) gives 
\be 
\oint m {\dot{x}}^2 {\rm d} t = 2 \pi m \sum_{\al = - \infty}^{\infty} \left|X_\al(n)\right|^2 \al^2 \omega(n)
\label{eq:2.14}
\ee
using $X_\al(n) = [X_{-\al}(n)]^*$; (\ref{eq:2.14}) is H(14). Heisenberg argues that (\ref{eq:2.14}) does not sit well with the Correspondence Principle, 
since the latter should only determine $J$ up to an additive constant (times 
$h$). Setting (\ref{eq:2.14}) equal to $n h$, he converts it to the form 
(H(15)) 
\be 
h = 2 \pi m \sum_{\al = - \infty}^{\infty} \al \frac{{\rm d}}{{\rm d} n} 
( \al \left| X_\al(n)\right|^2 \omega(n)), 
\label{eq:2.15}
\ee
which determines the $X_\al(n)$'s only to within a constant; the summation can 
alternatively be written as over positive values of $\al$, replacing 
$2 \pi m$ by $4 \pi m$. In another crucial jump, Heisenberg now replaces 
the differential in (\ref{eq:2.15}) by a {\em difference}, giving 
\be 
h = 4 \pi m \sum_{\al =0}^{\infty} \{ \left| X(n+\al, n)\right|^2\omega(n+\al, n) - \left|X(n, n-\al)\right|^2\omega(n, n-\al) \},
\label{eq:2.16}
\ee
which is H(16)$^{27}$. As he later recalled, he had noticed that `if I 
wrote down this [presumably (\ref{eq:2.15}) above] and tried to translate 
it according to the scheme of dispersion theory, I got the Thomas-Kuhn sum 
rule [which is equation (\ref{eq:2.16})$^{28, 29}$]. And that is the point. 
Then I thought, ``That is apparently how it is done'' '$^{30}$. 

By `the scheme of dispersion theory', Heisenberg is referring to what 
Jammer$^{31}$ calls Born's correspondence rule, namely$^{32}$ 
\be 
\al \frac{\partial \Phi(n)}{\partial n} \leftrightarrow \Phi(n) - \Phi(n-\al), 
\label{eq:2.17}
\ee
or rather to its iteration in the form$^{33}$
\be \al \frac{\partial \Phi(n, \al)}{\partial n} \leftrightarrow \Phi(n+\al, n) - \Phi(n, n-\al)
\label{eq:2.18}
\ee
as used in the Kramers-Heisenberg theory of dispersion$^{34, 35}$. It took 
Born only a few days to show that Heisenberg's quantum condition (\ref{eq:2.16}) was in fact the diagonal matrix element of (\ref{eq:2.11}), and to guess$^{36}$ 
that the off-diagonal elements of $\hat{x} \hat{p} - \hat{p} \hat{x}$ were 
zero, a result which was shown to be compatible with the equations of motion 
in Born and Jordan's paper$^4$. 

At this point, when we have momentarily advanced a little beyond July 1925, it 
may be appropriate to emphasize that Heisenberg's transition amplitude $X(n, n-\al)$ is indeed precisely the same as the quantum-mechanical matrix 
element $\langle n-\al|\hat{x}|n\rangle$, where $|n\rangle$ is the exact 
eigenstate with energy $W(n)$. The relation of (\ref{eq:2.16}) to the 
fundamental commutator (\ref{eq:2.11}) is briefly recalled       
 in Appendix A. 

Returning to the development of Heisenberg's paper, he summarizes the state of 
affairs reached so far by the statement that  equations (\ref{eq:2.12}) and 
(\ref{eq:2.16}) `if soluble,  contain a complete determination not only of the frequencies and energies, but also of the 
quantum theoretical transition probabilities'. We draw attention to the strong 
claim here: that he has arrived at a new {\em calculational scheme}, which 
will {\em completely determine} the {\em observable quantities}. Let us now 
see in detail how this scheme works, for the case of a harmonic oscillator 
perturbed by an anharmonic force of the form $\lambda x^2$ per unit mass$^{37}$. 

\section{ Heisenberg's calculational scheme, and its detailed working-out for  the `$\lambda x^2$' anharmonic oscillator}

\subsection{Recursion relations in the quantum case}

The classical (and, by assumption, quantum-mechanical) equation of motion is 
\be 
\ddot{x} + \omega_0^2 x + \lambda x^2 =0.
\label{eq:2.19}
\ee
Departing now from the order of Heisenberg's presentation, we shall begin by 
showing how - as he states   - (\ref{eq:2.19}) leads to  recursion relations for the transition amplitudes $X(n, n-\al)$. The $(n, n-\al)$ 
representative of the first two terms in (\ref{eq:2.19}) is 
straightforward, being just 
\be 
[-\omega^2(n, n-\al) + \omega^2_0] X(n, n-\al) {\rm e}^{{\rm i} \omega(n, n-\al) t}, 
\label{eq:2.20}
\ee
while that of the third term is, by the rule (\ref{eq:2.10}), 
\be 
\lambda \sum_\beta X(n, n-\beta) X(n-\beta, n-\al) {\rm e}^{{\rm i} \omega(n, n-\al) t}.
\label{eq:2.21}
\ee
The $(n, n-\al)$ representative$^{38}$ of (\ref{eq:2.19}) therefore yields$^{39}$
\be 
[\omega^2_0 - \omega^2(n, n-\al)] X(n, n-\al) + \lambda \sum_{\beta} X(n, n-\beta) X(n-\beta, n-\al) =0, 
\label{eq:2.22}
\ee
which generates a recursion relation for each value of $\al$ ($\al = 0, \pm 1, \pm 2, \ldots $). For example, setting $\al=0$ we obtain 
\be 
\omega^2_0 X(n, n) + \lambda [ X(n, n) X(n, n) + X(n, n-1) X(n-1, n) + X(n, n+1) X(n+1, n) + \ldots].
\label{eq:2.23}
\ee
No general solution for this infinite set of non-linear algebraic equations 
seems to be possible, so - following Heisenberg - we turn to a perturbative 
approach. 

\subsection{Perturbation theory}

To make the presentation self-contained, we need to include a certain 
number of ancillary results. Heisenberg begins by considering the perturbative 
solution of the classical equation (\ref{eq:2.12}). He writes the solution in 
the form 
\be 
x(t) = \lambda a_0 + a_1 \cos \omega t + \lambda a_2 \cos 2 \omega t + \lambda^2 a_3 \cos 3 \omega t + \ldots + \lambda^{\al -1} a_{\al} \cos \al \omega t + \ldots 
\label{eq:2.24}
\ee
where the coefficients $a_{\al}$, and $\omega$, are themselves to be expanded as a power series in $\lambda$, the first terms of which are independent of $\lambda^{38}$:
\bea
a_0 &=& a^{(0)}_0 + \lambda a^{(1)}_0 + \lambda^2 a^{(2)}_0 + \ldots \label{eq:2.25}\\
a_1  &=& a^{(0)}_1 + \lambda a^{(1)}_1 + \lambda^2 a^{(2)}_1 + \ldots \label{eq:2.26}
\eea
and so on, and 
\be 
\omega = \omega_0 + \lambda \omega^{(1)} + \lambda^2 \omega^{(2)} + \ldots . \label{eq:2.27}
\ee
 Straightforwardly inserting (\ref{eq:2.24}) into (\ref{eq:2.12}), using 
standard trigonometric identities, and equating to zero the terms 
which are constant, and which multiply $\cos \omega t$, $\cos 2 \omega t$, 
etc, one obtains 
\be
({\rm constant}) \ \ \ \lambda \{\omega^2_0 a_0 + \frac{1}{2}a^2_1 + [\lambda^2(a^2_0+\frac{1}{2}a^2_2) + \ldots]\} = 0 \label{eq:2.28}
\ee
\be
(\cos \omega t) \ \ \  (- \omega^2 + \omega^2_0)a_1 + [ \lambda^2 ( a_1a_2 + 2 a_0a_1) + \ldots ] =0 \label{eq:2.29} 
\ee
\be 
(\cos 2 \omega t) \ \ \ \lambda\{(-4 \omega^2 + \omega^2_0)a_2 + \frac{1}{2}a_1^2 + [\lambda^2(a_1a_3 + 2 a_0a_2) + \ldots ]\}=0 \label{eq:2.30}
\ee
\be 
(\cos 3 \omega t) \ \ \  \lambda^2\{(-9\omega^2 + \omega^2_0)a_3 + a_1a_2 + [ \lambda^2(a_1a_4 + 2 a_0a_3) + \ldots]\}=0 \label{eq:2.31}
\ee
and so on, where the dots stand for higher powers of $\lambda$. Dropping all 
the terms multiplying $\lambda^2$ (and higher powers), (\ref{eq:2.28}) - (\ref{eq:2.31}) become (for $\lambda \neq 0$ and $a_1 \neq 0$) 
\bea
\omega^2_0 a_0 + \frac{1}{2} a^2_1 & = & 0 \label{eq:2.32} \\
(-\omega^2 + \omega^2_0) & = & 0 \label{eq:2.33} \\
(-4\omega^2 + \omega^2_0) a_2 + \frac{1}{2}a^2_1 &=&0 \label{eq:2.34}\\
(-9\omega^2 + \omega^2_0) a_3 + a_1a_2 &=&0 \label{eq:2.35} 
\eea
which are the same as H(18)$^{41}$. The `lowest order in $\lambda$' solution 
is now obtained from (\ref{eq:2.32}) - (\ref{eq:2.35}) by 
setting $\omega = \omega_0$, and replacing each $a_\al$ 
 by the corresponding one with a superscript `$\mbox{}^{(0)}$' (c.f. 
(\ref{eq:2.25}),  (\ref{eq:2.26})). 

Turning now to the quantum case, Heisenberg proposes to seek a solution 
analogous to (\ref{eq:2.24}). Of course, it is now a matter of using 
the `representation' of $x(t)$ in terms of the quantities $\{X(n, n-\al){\rm exp} [{\rm i} \omega(n, n-\al)t]\}.$ But it seems reasonable to assume that, as 
the index $\al$ increases away from zero, in integer steps, each 
successive amplitude 
will (to leading order in $\lambda$) be suppressed by an additional power 
of $\lambda$, as in the classical case.  Thus Heisenberg suggests that, in the quantum case, $x(t)$ 
should be represented by terms of the form 
\bea 
&&\lambda a(n, n), \  a(n, n-1) \cos \omega(n, n-1) t, \ \lambda a(n, n-2) \cos \omega(n, n-2)t,  \ldots , \nonumber \\
 &&\lambda^{\al -1} a(n, n-\al) \cos \omega(n, n-\al)t, \ldots, 
\label{eq:2.36}
\eea
where, as in (\ref{eq:2.25}) - (\ref{eq:2.27}), 
\be 
a(n, n) = a^{(0)}(n, n) + \lambda a^{(1)}(n, n) + \lambda^2 a^{(2)}(n, n) + \ldots \label{eq:2.37}
\ee
\be 
 a(n, n-1) = a^{(0)}(n, n-1) + \lambda a^{(1)}(n, n-1) + \lambda^2 a^{(2)}(n, n-1) + \ldots \label{eq:2.38}
\ee
and so on, and 
\be 
\omega(n, n-\al) = \omega^{(0)}(n, n-\al) + \lambda \omega^{(1)}(n, n-\al) + 
\lambda^2 \omega^{(2)}(n, n-\al) + \ldots \label{eq:2.39}
\ee
As Born and Jordan pointed out$^{4}$, some use of `correspondence' arguments 
has been made here, in assuming that,  as $\lambda \to 0$, only transitions 
between adjacent states are possible (we shall return to this point in 
section 3.3). 

Heisenberg now simply {\em writes down} what he asserts to be the quantum 
versions of (\ref{eq:2.32}) - (\ref{eq:2.35}), namely$^{42}$ 
\be 
\omega^2_0 a(n, n) + \frac{1}{4}[a^2(n+1, n) + a^2(n, n-1)]=0 \label{eq:2.40}
\ee
\be 
- \omega^2(n, n-1) + \omega^2_0 = 0 \label{eq:2.41}
\ee
\be 
[-\omega^2(n, n-2) + \omega^2_0]a(n, n-2) + \frac{1}{2} [a(n, n-1)a(n-1, n-2)] = 0 \label{eq:2.42}
\ee
\bea 
&&[- \omega^2(n, n-3) + \omega^2_0] a(n, n-3) + \frac{1}{2}[a(n, n-1)a(n-1, n-3) \nonumber \\
&& + a(n, n-2)a(n-2, n-3)] = 0 .\label{eq:2.43}
\eea
The question we now want to address  is: 
how did Heisenberg arrive at equations (\ref{eq:2.40}) - (\ref{eq:2.43})? 

Tomonaga$^{8}$, having derived (\ref{eq:2.22}), then proceeds to discuss only 
the $\lambda \to 0$ limit - i.e. the simple harmonic oscillator, a special case 
to which we shall return. The only other authors, to our knowledge, who have 
offered a discussion of the presumed details of Heisenberg's calculations 
are$^{43}$ Mehra and Rechenberg$^{11}$. 
They suggest that Heisenberg guessed how to 
`translate', `re-interpret'  or `reformulate' (their words) the classical equations 
(\ref{eq:2.32}) - (\ref{eq:2.35}) into the quantum ones (\ref{eq:2.40}) - (\ref{eq:2.43}), in 
a way that was consistent with his multiplication law (\ref{eq:2.10}). Although 
such `inspired guesswork' was undoubtedly necessary in the stages leading up 
to Heisenberg's paper of July 1925, to us it seems more 
plausible that by the time 
of the paper's final formulation, Heisenberg realised that he had a 
calculational scheme in which guesswork was no longer necessary, and in 
which (\ref{eq:2.40}) - (\ref{eq:2.43}), in particular, could be derived.

Unfortunately, we know of  no documentary evidence which can directly 
prove (or disprove) this suggestion. But we think there is some internal 
evidence for it. In the passage to which attention 
was drawn earlier, Heisenberg 
firmly asserts that the formalism he has set up constitutes a complete 
scheme for calculating everything that needs to be calculated. It is hard to 
believe that Heisenberg did not realise that it led directly to equations 
(\ref{eq:2.40}) - (\ref{eq:2.43}), without the need for any `translations' of the 
classical relations (of course, the latter were a nice check on the reasonableness 
of the quantum relations). At all events, it is the case that (\ref{eq:2.40}) - 
(\ref{eq:2.43}) can be straightforwardly derived, as we shall  now see. 

In order to apply the ansatz (\ref{eq:2.36}) to (\ref{eq:2.22}), we need to 
relate the amplitudes $X(n, n-\al)$ to the corresponding quantities 
$\lambda^{\al -1} a(n, n-\al)$. We first note that in the classical case
\be 
X_{\al}(n) = X^*_{-\al}(n) \label{eq:3.1}   
\ee
since $x(t)$ in (\ref{eq:2.4}) has to be real. Consider, without loss of generality, the case $\alpha > 0$. Then the quantum-theoretical analogue of 
the left hand side of (\ref{eq:3.1}) is $X(n, n-\alpha)$, while that of the 
right hand side is $X^{*}(n-\alpha, n)$ (see Ref. 24). Hence  the quantum-theoretical analogue 
of (\ref{eq:3.1}) is 
\be
X(n, n-\al) = X^*(n - \al, n), \label{eq:3.2}
\ee
which of course is nothing but the relation $\langle n-\al|\hat{x}|n\rangle = 
\langle n | \hat{x} | n  -\al \rangle^*$ for the Hermitian observable 
$\hat{x}$. Although $X(n, n-\al)$ can in principle be complex (and he twice 
discusses the significance of the phases of such amplitudes), Heisenberg 
seems to have assumed (as is certainly plausible) that in the context of the 
classical cosine expansion (\ref{eq:2.24}), and the corresponding quantum terms (\ref{eq:2.36}), 
 the $X(n, n-\al)$'s should  be chosen 
to be real, so that (\ref{eq:3.2}) becomes 
\be 
X(n, n-\al) = X(n-\al, n) \label{eq:3.3}
\ee
- that is, the matrix with elements $\{X(n, n-\al)\}$ is symmetric. Consider 
now a typical term of (\ref{eq:2.36})
\bea 
&&\lambda^{\al -1} a(n, n-\al) \cos [ \omega(n, n-\al) t ]  \nonumber \\
&=& \frac{\lambda^{\al -1}}{2} a(n, n-\al)\{{\rm exp}[{\rm i}\omega(n, n-\al)t] + {\rm exp}[-{\rm i} \omega(n, n-\al)t]\} =\nonumber\\
&=& \frac{\lambda^{\al-1}}{2}a(n, n-\al) \{{\rm exp}[{\rm i}\omega(n, n-\al)t]+
 {\rm exp}[{\rm i}\omega(n-\al, n) t]\} 
\label{eq:3.4}
\eea
using $\omega(n, n-\al) = - \omega(n-\al, n)$ from (\ref{eq:2.1}). Assuming 
that $a(n, n-\al) = a(n-\al, n)$ as discussed for (\ref{eq:3.3}), we see that 
it is consistent to write 
\be 
X(n, n-\al) = \frac{\lambda^{\al-1}}{2}a(n, n-\al) \label{eq:3.5}
\ee
for positive values of $\alpha$, and in general 
\be
X(n, n-\al) = \frac{\lambda^{|\al|-1}}{2} a(n, n-\al), \ \ \ \al \neq 0. \label{eq:3.6}
\ee
The particular case $\al=0$ is clearly special, with $X(n, n)=\lambda a(n, n)$. 

We may now write out the recurrence relations (\ref{eq:2.22}) explicitly for 
 $\al = 0, 1, 2, \ldots$, in terms of the $a(n, n-\al)$ 
rather than the $X(n, n-\al)$. We shall include terms up to and including 
$\lambda^2$. 

For $\al = 0$ we obtain
\bea
&&\lambda \{ \omega^2_0 a(n, n) + \frac{1}{4}[a^2(n+1, n) + a^2(n, n-1)] + \nonumber \\
&&\lambda^2[a^2(n, n) + \frac{1}{4}(a^2(n+2, n) + a^2(n, n-2))]\}=0. \label{eq:3.7}
\eea
 We note the connection with (\ref{eq:2.28}), and that (\ref{eq:3.7}) reduces to 
Heisenberg's (\ref{eq:2.40}) when the $\lambda^2$ term is dropped. Similarly, for 
$\al =1$ we obtain 
\bea
&&(-\omega^2(n, n-1) + \omega^2_0)a(n, n-1) + \lambda^2\{a(n, n)a(n, n-1) + 
a(n, n-1)a(n-1, n-1) + \nonumber \\
&&\frac{1}{2}[a(n, n+1)a(n+1, n-1) + a(n, n-2)a(n-2, n-1)]\}=0 \label{eq:3.8} 
\eea
(c.f. (\ref{eq:2.29})); for $\al = 2$ 
\bea
&&\lambda\{(-\omega^2(n, n-2) + \omega^2_0)a(n, n-2) + \frac{1}{2}a(n, n-1)a(n-1, n-2) + \nonumber\\
&&\lambda^2[a(n, n)a(n, n-2) + a(n, n-2)a(n-2, n-2) + \frac{1}{2}a(n, n+1)a(n+1, n-2) \nonumber \\
&&+\frac{1}{2}a(n, n-3)a(n-3, n-2)]\}=0 \label{eq:3.9}
\eea
(c.f. (\ref{eq:2.30})); and for $\al =3$ (c.f. (\ref{eq:2.31})) 
\bea 
&&\lambda^2\{(-\omega^2(n, n-3) + \omega^2_0)a(n, n-3) + \nonumber \\
&&\frac{1}{2}[a(n, n-1)a(n-1, n-3) + a(n, n-2)a(n-2, n-3)] +\nonumber \\
&& \lambda^2[a(n, n)a(n, n-3) + a(n, n-3) a(n-3, n-3) + \frac{1}{2}a(n, n+1)a(n+1, n-3) + \nonumber \\
&&\frac{1}{2}a(n, n-4)a(n-4, n-3)]\} = 0 . \label{eq:3.10}
\eea
On dropping the terms multiplied by $\lambda^2$, (\ref{eq:3.7}) - (\ref{eq:3.10}) reduce to Heisenberg's (\ref{eq:2.40}) - (\ref{eq:2.43}). This 
appears to be the first published derivation of these equations. 

In addition to these recurrence relations which follow from the equations of 
motion, we also need the perturbative version of the quantum condition 
(\ref{eq:2.16})$^{44}$. Including terms with a $\lambda^2$ power, consistent with 
(\ref{eq:3.7}) - (\ref{eq:3.10}), (\ref{eq:2.16}) becomes
\bea
&&\frac{h}{\pi m} = a^2(n+1, n) \omega(n+1, n) - a^2(n, n-1) \omega(n, n-1) \nonumber \\
 && + \lambda^2 [ a^2(n+2, n)\omega(n+2, n) - a^2(n, n-2)\omega(n, n-2)]. \label{eq:3.11}
\eea
We are now ready to obtain the solutions. 

\subsection{The lowest-order solutions for the amplitudes and frequencies}  

We begin by considering the lowest-order solutions, in which all $\lambda^2$ 
terms are dropped from equations (\ref{eq:3.7}) - (\ref{eq:3.11}), and all quantities 
($a$'s and $\omega$'s) are replaced by the corresponding ones with a superscript $\mbox{}^{(0)}$ (c.f. (\ref{eq:2.37}) - (\ref{eq:2.39})$^{45}$. In this case, (\ref{eq:3.8}) reduces to 
\be
[-(\omega^{(0)}(n, n-1))^2 + \omega^2_0]a^{(0)}(n, n-1)=0, \label{eq:3.12}
\ee
so that assuming $a^{(0)}(n, n-1) \neq 0$ we obtain 
\be 
\omega^{(0)}(n, n-1) = \omega_0
\label{eq:3.13}
\ee
for all $n$. Substituting (\ref{eq:3.13}) into the lowest-order version of (\ref{eq:3.11}) we find 
\be
\frac{h}{\pi m \omega_0} = [a^{(0)}(n+1, n)]^2 - [a^{(0)}(n, n-1)]^2. \label{eq:3.14}
\ee
The solution of this difference equation is 
\be 
[a^{(0)}(n, n-1)]^2 = \frac{h}{\pi m \omega_0}(n + {\rm constant}), \label{eq:3.15}
\ee
as given in H(20)$^{45}$. To determine the value of the constant, Heisenberg 
used the idea that in the ground state there can be no transition to a lower 
state. Thus 
\be 
[a^{(0)}(0, -1)]^2 =0 \label{eq:3.16}
\ee
and the constant is determined to be zero. This then gives (up to a convention 
as to sign) 
\be 
a^{(0)}(n, n-1) = \beta \sqrt{n} \label{eq:3.17}
\ee
where 
\be 
\beta = (h/\pi m \omega_0)^{1/2}. \label{eq:3.18}
\ee

Equations (\ref{eq:3.13}) and (\ref{eq:3.17}) are Heisenberg's first results, and they 
are in fact appropriate to the simple (unperturbed) oscillator. We can check 
(\ref{eq:3.17}) against standard quantum mechanics via 
\be 
a^{(0)}(n, n-1)=2 X^{(0)}(n, n-1) = 2 \; _0\langle n-1 | \hat{x} | n \rangle_0 \label{eq:3.19}
\ee
where the states $| n \rangle_0$ are unperturbed oscillator eigenstates. It is well 
known that $^{46}$ 
\be 
_0\langle n-1 | \hat{x} | n \rangle_0 = \left( \frac{\hbar}{2 m \omega_0}\right)^{1/2} \sqrt{n}
\label{eq:3.20}
\ee
which agrees with (\ref{eq:3.17}), using (\ref{eq:3.18}). 
A similar treatment of (\ref{eq:3.7}) leads to 
\be 
a^{(0)}(n, n) = - \frac{\beta^2}{4 \omega^2_0}(2n+1). \label{eq:3.21}
\ee

Turning next to equation (\ref{eq:3.9}), the lowest-order form is 

\be
(-[\omega^{(0)}(n, n-2)]^2+\omega^2_0)a^{(0)}(n, n-2) + \frac{1}{2}a^{(0)}(n, n-1)a^{(0)}(n-1, n-2) = 0. \label{eq:3.22}
\ee
Now the composition law must be true for the lowest-order frequencies, so we 
have 
\bea
\omega^{(0)}(n, n-2)&=&\omega^{(0)}(n, n-1) + \omega^{(0)}(n-1, n-2) \nonumber \\
&=& 2 \omega_0 \label{eq:3.23}
\eea
using (\ref{eq:3.13}), and in general 
\be
\omega^{(0)}(n, n-\al) = \al \omega_0 \ \ \ (\al = 1, 2, 3, \ldots). \label{eq:3.24}
\ee
Using (\ref{eq:3.17}), (\ref{eq:3.23}) and (\ref{eq:3.24}) we then obtain 
\be 
a^{(0)}(n, n-2) = \frac{\beta^2}{6 \omega^2_0} \sqrt{n(n-1)}. \label{eq:3.25}
\ee
 A similar teatment of (\ref{eq:3.10}) yields 
\be a^{(0)}(n, n-3) = \frac{\beta^3}{48 \omega^4_0} \sqrt{n(n-1)(n-2)}. \label{eq:3.26}
\ee
Consideration of the general lowest-order term in (\ref{eq:2.22}) shows that 
\be 
a^{(0)}(n, n-\al) = A_{\al} \frac{\beta^{\al}}{\omega_{0}^{2(\al-1)}} \sqrt \frac{n!}
{(n-\al)!}, \label{eq:3.27}
\ee
where $A_{\al}$ is a numerical factor depending on $\al$; (\ref{eq:3.27}) is  equivalent to H(21).

At this stage it is instructive to comment briefly on the relation of the 
above results to those which would be obtained in standard quantum-mechanical 
perturbation theory. At first sight, it is surprising to see non-zero 
amplitudes for two-quantum ((\ref{eq:3.25})), three-quantum 
((\ref{eq:3.26})) or 
$\al$-quantum ((\ref{eq:3.27})) transitions appearing at `lowest order'. But we have to 
remember that in Heisenberg's perturbative ansatz (\ref{eq:2.36}), the 
$\al$-quantum amplitude appears multiplied by a factor $\lambda^{\al-1}$. 
Thus, for example, the `lowest order' two-quantum amplitude is really 
$\lambda a^{(0)}(n, n-2)$, not just $a^{(0)}(n, n-2)$. Indeed, such a 
transition is to be expected precisely at order $\lambda^1$, in conventional 
perturbation theory. The amplitude is $\langle n-2 | \hat{x} | n \rangle$ 
where, to order $\lambda$, 
\be
|n\rangle = | n \rangle_0 + \frac{1}{3} m \lambda \sum_{k \neq n} \frac{_0\langle k | {\hat{x}}^3 | n \rangle_0}{(n-k) \hbar \omega_0} | k \rangle_0. \label{eq:3.28}
\ee
The operator ${\hat{x}}^3$ connects $| n \rangle_0$ to $| n+3 \rangle_0, | n + 1 \rangle_0, | n-1 \rangle_0$, and $| n-3 \rangle_0$, and similar connections 
occur for $_0\langle  n-2 |$, so that a non-zero $O(\lambda)$ amplitude is 
generated in $\langle n-2 | \hat{x} | n \rangle$. 

It is straightforward to check that (\ref{eq:3.25}) is indeed correct  quantum-mechanically; but it is more tedious to check (\ref{eq:3.26}), and distinctly unpromising to 
contemplate checking (\ref{eq:3.27}) by doing a conventional perturbation 
calculation to order $\al -1$. For this particular problem, the `improved' 
perturbation theory represented by (\ref{eq:2.36}) is clearly very useful. 

Having calculated the amplitudes for this problem to lowest order, Heisenberg 
next considers the energy. Unfortunately he again gives no details of his 
calculation, beyond saying that he uses the classical expression for the 
energy, namely 
\be 
W=\frac{1}{2}m{\dot{x}}^2 + \frac{1}{2}m\omega^2_0x^2 + \frac{1}{3}m \lambda x^3. \label{eq:3.29}
\ee
It seems a reasonable conjecture, however, that he replaced each term in (\ref{eq:3.29}) by its `ensemble of representatives'. Thus $x^2$, for 
example, is replaced by the ensemble of terms of the form 
\be
\sum_\beta X(n, n-\beta) X(n-\beta, n-\al) {\rm e}^{{\rm i} \omega(n, n-\al) t}, \label{eq:3.30}
\ee
according to his composition law (\ref{eq:2.10}). Similarly, ${\dot{x}}^2$ is replaced by terms of the form 
\bea
&&\sum_\beta{\rm i}\omega(n, n-\beta)X(n, n-\beta){\rm e}^{{\rm i}\omega(n, n-\beta)t} 
{\rm i} \omega(n-\beta, n-\al)X(n-\beta, n-\al){\rm e}^{{\rm i} \omega(n-\beta, n-\al)t} \nonumber \\
&&= \sum_\beta \omega(n, n-\beta)\omega(n-\al, n-\beta)X(n, n-\beta)X(n-\beta, n-\al) {\rm e}^{{\rm i}\omega(n, n-\al)t}, \label{eq:3.31}
\eea
using $\omega(n, m) = - \omega(m, n)$. In general then, in Heisenberg's 
scheme, the `energy representatives' will be of the form 
\be 
W(n, n-\al) {\rm e}^{{\rm i}\omega(n, n-\al)t}. \label{eq:3.32}
\ee
It follows that if energy is to be conserved (i.e.\,time-independent) the 
{\em off-diagonal elements must vanish}:
\be 
W(n, n-\al)=0, \ \ \ \al \neq 0. \label{eq:3.33}
\ee
The term $\al=0$ is time-independent, and may be 
taken to be the energy in the state 
$n$. The crucial importance of checking the condition (\ref{eq:3.33}) was 
clearly appreciated by Heisenberg. 

To lowest order in $\lambda$, the last term in (\ref{eq:3.29}) may be dropped. Furthermore, 
referring to (\ref{eq:2.36}), the only $\lambda$-independent terms in 
the $X$-amplitudes are those involving one-quantum jumps such as 
$n \to n-1 $, corresponding in lowest order to amplitudes such as 
$X^{(0)}(n, n-1) = \frac{1}{2}a^{(0)}(n, n-1)$. This means 
 (referring to (\ref{eq:3.30}) and (\ref{eq:3.31})) that the elements
$W(n, n), W(n, n-2)$ and $W(n, n+2)$ , and only these $W(n, m)$'s, are 
 independent of $\lambda$ when 
evaluated to lowest order. In Appendix B we show that $W(n, n-2)$ vanishes 
to lowest order, and $W(n, n+2)$ vanishes similarly. 
Thus, to lowest order in $\lambda$, the energy is indeed conserved (as Heisenberg notes), and is given (using (\ref{eq:3.30}) and (\ref{eq:3.31}) with  $\al=0$ and $\beta = \pm 1$) by  
\bea
W(n, n) &=& \frac{1}{2}m[\omega^{(0)}(n, n-1)]^2[X^{(0)}(n, n-1)]^2 + \nonumber \\
&& \frac{1}{2}m 
[\omega^{(0)}(n+1, n)]^2[X^{(0)}(n+1, n)]^2 \nonumber \\
&& + \frac{1}{2}m\omega^2_0[X^{(0)}(n, n-1)]^2 + \frac{1}{2}m \omega^2_0[X^{(0)}(n+1, n)]^2 \nonumber \\
&=&(n+\frac{1}{2}) \hbar \omega_0 \label{eq:3.34}
\eea
using (\ref{eq:3.13}), (\ref{eq:3.17}) and (\ref{eq:3.18}). This is the result given 
by Heisenberg in H(23). 

These `lowest order' results are the only ones Heisenberg reports for the $\lambda x^2$ force. We do not know whether he carried out higher-order 
calculations for this case or not. What he writes, next, is that the `more 
precise calculation, taking into account higher order approximations in $W, a$ 
and $\omega$ will now be carried out for the simpler example of an 
anharmonic oscillator $\ddot{x} + \omega^2_0 + \lambda x^3 = 0$ .' This case is 
slightly simpler because only the `odd' terms are present in (\ref{eq:2.36}) 
(i.e. $a_1, \lambda a_3, \lambda^2 a_5$, etc). The results Heisenberg 
states for the `$\lambda x^3$' problem include terms up to order $\lambda$ 
in the amplitudes, and terms up to order $\lambda^2$ in the frequency $\omega(n, n-1)$, and in the energy $W$. Once again, he gives no details of how he has 
done the calculations. We believe there can be little doubt that he went 
through the algebra of solving the appropriate recurrence relations, up to 
order $\lambda^2$ in the requisite quantities. As far as we know, the details of such a calculation have not been given before, and we therefore feel that it 
is worth giving them here, as being of both pedagogical 
and historical interest. In the 
following section we shall obtain the solutions for the $\lambda x^2$ force 
(up to order $\lambda^2$) which we have been considering hitherto, rather than 
start afresh with the $\lambda x^3$ one. The procedure is of course the same 
for both. 

Before leaving the `lowest order' calculations, we address a question which 
may have occurred to the reader: given that, at this stage in his paper, 
the main results actually relate to the simple harmonic oscillator rather 
than to the anharmonic one, why did Heisenberg not begin his discussion 
of `toy models' with the simplest one of all, namely the simple harmonic 
oscillator? And indeed, is it not possible to apply his procedure to the 
SHO, without going through the apparent device of introducing a perturbation, 
and then retaining only those parts of the solution which survive as the 
perturbation vanishes? 

For the unperturbed oscillator (the SHO), the equation of motion is of 
course $\ddot{x} + \omega^2_0 x= 0$, which yields 
\be 
[\omega^2_0 - \omega^2(n, n-\al)] X(n, n-\al) = 0 \label{eq:3.35}
\ee
for the amplitudes $X$ and frequencies $\omega$. It is also reasonable to 
retain the `quantum condition' (\ref{eq:2.16}), since this is supposed to be true 
whatever the particular force law. If we {\em assume} that the only non-vanishing amplitudes are those involving adjacent states (because, 
for example,  in the classical case
 only a single harmonic is present$^{47}$), then - remembering that $X(n, n-1) = \frac{1}{2} a(n, n-1)$ - 
(\ref{eq:2.16}) and (\ref{eq:3.35}) reduce to (\ref{eq:3.14}) and (\ref{eq:3.12}) respectively, and we quickly recover the previous results. Thus 
we see that this is indeed an efficient way 
to solve the quantum SHO$^{48}$. For completeness, however, it would be 
nice not to have to make the `adjacent states' assumption; Born and 
Jordan$^{4}$ showed how this could be done, but their argument is 
somewhat involved. Soon thereafter, of course, the wave-mechanics of 
Schr\"{o}dinger, and the operator approach of Dirac, provided the derivations 
used ever since. 

We now turn to the higher order corrections, for the $\lambda x^2$ force. 

\subsection{The solutions up to and including $\lambda^2$ terms}

Consider first equation (\ref{eq:3.8}), retaining terms of order $\lambda$ but 
no higher powers. We set 
\bea
\omega(n, n-1) &=& \omega_0 + \lambda \omega^{(1)}(n, n-1) \label{eq:3.36}\\
a(n, n-1) &=& a^{(0)}(n, n-1) + \lambda a^{(1)}(n, n-1) \label{eq:3.37}
\eea
and find 
\be 
2 \lambda \omega_0 \omega^{(1)}(n, n-1) a^{(0)}(n, n-1) = 0, \label{eq:3.38}
\ee
so that 
\be 
\omega^{(1)}(n, n-1) =0. \label{eq:3.39}
\ee
Consideration of equation (\ref{eq:3.8}) up to terms of order $\lambda^2$, employing equations (\ref{eq:3.17}), (\ref{eq:3.21}) and (\ref{eq:3.25}) for the 
zeroth order amplitudes, gives the $O(\lambda^2)$ correction to $\omega(n, n-1)$ 
(c.f. (\ref{eq:2.27})):
\be 
\omega^{(2)}(n, n-1) = -\frac{ 5 \beta^2}{12 \omega^3_0} n. \label{eq:3.40}
\ee

The corresponding corrections to $a(n, n-1)$ are found from the quantum condition (\ref{eq:2.16}). To order $\lambda$ we set 
\be
a(n+1, n) = a^{(0)}(n+1, n) + \lambda a^{(1)}(n+1, n) \label{eq:3.41}
\ee
as in (\ref{eq:3.37}), and find 
\be 
\sqrt{n+1} a^{(1)}(n+1, n) - \sqrt{n} a^{(1)} (n, n-1) =0. \label{eq:3.43}
\ee
This equation has the solution $a^{(1)}(n, n-1) = {\rm constant} /\sqrt{n}$, but the 
condition $a^{(1)}(0, -1) = 0$ (c.f. (\ref{eq:3.16})) implies that the constant 
must be zero, and so   
\be 
a^{(1)}(n, n-1) = 0. \label{eq:3.44}
\ee
In a similar way, to order $\lambda^2$ we obtain 
\be      
\sqrt{n+1}a^{(2)}(n+1, n) - \sqrt{n}a^{(2)}(n, n-1) = \frac{11 \beta^3}{72 \omega^4_0} (2n+1), \label{eq:3.45}
\ee
which has the solution 
\be 
a^{(2)}(n, n-1) = \frac{11 \beta^3}{72 \omega^4_0} n \sqrt{n}. \label{eq:3.46}
\ee

We can now find the higher order corrections to $a(n, n)$ by considering equation 
(\ref{eq:3.7}). We obtain 
\be 
a^{(1)}(n, n) =0, \ \ \ \ a^{(2)}(n, n)= - \frac{\beta^4}{72 \omega^6_0}(30n^2+30n+11). \label{eq:3.47}
\ee
Similarly, we find from (\ref{eq:3.9}) 
\be 
a^{(1)}(n, n-2)=0, \ \ \ \ a^{(2)}(n, n-2) = \frac{3 \beta^4}{32 \omega^6_0}(2n-1)\sqrt{n(n-1)}, \label{eq:3.48}
\ee
where we have used 
\bea
\omega^{(2)}(n, n-2) &=& \omega^{(2)}(n, n-1) + \omega^{(2)}(n-1, n-2) \nonumber \\
&=& - \frac{5 \beta^2}{12 \omega^3_0} (2n-1). \label{eq:3.49}
\eea
These results suffice for our purpose. When $n$ is large, they agree with those 
obtained for the classical `$\lambda x^2$' anharmonic oscillator using the method of 
successive approximations$^{49}$. As an indirect check of their quantum-mechanical 
validity, we now turn to the energy, evaluated to order $\lambda^2$.

Consider first the $(n, n)$ element of $\frac{1}{2}m \omega^2_0 \hat{x}^2$. This is given, to order $\lambda^2$, by 
\bea
&&\frac{1}{2}m \omega^2_0\{\frac{1}{4}[(a^{(0)}(n, n-1))^2 + (a^{(0)}(n, n+1))^2]+  
\frac{\lambda^2}{4}[4(a^{(0)}(n, n))^2 + \nonumber \\ 
&&2a^{(2)}(n, n-1)a^{(0)}(n-1, n) + 
2a^{(2)}(n, n+1)a^{(0)}(n+1, n) + \nonumber \\
&&(a^{(0)}(n, n-2))^2 + (a^{(0)}(n, n+2))^2]\} \nonumber \\
&& = \frac{1}{2}m \omega^2_0 \{ \frac{\beta^2}{2}(n+\frac{1}{2}) + \frac{5 \beta^4 \lambda^2}{12 \omega^4_0}(n^2 + n + 11/30)\}. \label{eq:3.50}
\eea
Similarly, using (\ref{eq:3.31}) up to order $\lambda^2$, with $\al=0$, the 
$(n, n)$ element of $\frac{1}{2}m {\dot{\hat{x}}}^2$ is found to be 
\be
\frac{1}{2}m\omega^2_0\{\frac{\beta^2}{2}(n+\frac{1}{2}) - \frac{5 \beta^4 \lambda^2}{24 \omega^4_0} (n^2 + n + 11/30)\}. \label{eq:3.51} 
\ee
Finally we need to consider the $(n, n)$ element of the potential 
energy $\frac{1}{3}m \lambda {\hat{x}}^3$. To obtain the result to order 
$\lambda^2$, we need compute the $(n, n)$ element of ${\hat{x}}^3$ only 
to order $\lambda$. Using 
\be 
{\hat{x}}^3(n, n) = \sum_\al \sum_\beta X(n, n-\al) X(n-\al, n-\beta) X(n-\beta, n) \label{eq:3.52}
\ee
one finds that there are no zeroth order terms, but twelve terms of order $\lambda$ (recall that amplitudes such as $X(n, n)$ and $X(n, n-2)$ each carry 
one power of $\lambda$). Evaluating these terms using the previously obtained 
results gives 
\be  
 - \frac{5 m \lambda^2 \beta^4}{24 \omega^2_0} (n^2 + n + 11/30) \label{eq:3.53}
\ee
for this term in the energy. Combining (\ref{eq:3.50}), (\ref{eq:3.51}) and (\ref{eq:3.53}) 
then gives, for the energy up to order $\lambda^2$, 
\be 
W(n, n) = (n+\frac{1}{2})\hbar \omega_0 - \frac{5 \lambda^2 \hbar^2}{12 m \omega^4_0}(n^2 + n +11/30), \label{eq:3.54}
\ee
a result$^{50}$ which agrees with classical perturbation theory when $n$ is 
large$^{51}$, and is in exact agreement with standard second-order perturbation theory in quantum mechanics$^{52}$. 

As mentioned earlier, Heisenberg does not give results for the `$\lambda x^2$' force beyond zeroth order; he does, however, give the results for the `$\lambda x^3$' force up to and including $\lambda^2$ terms in the energy, and $\lambda$ terms in the amplitudes. By `the energy' we mean, as usual, the $(n, n)$ element of the energy operator, which as noted in section 3.3 is independent of time. One should also check that the off-diagonal elements $W(n, n-\al)$ vanish (see equation (\ref{eq:3.33})). These are the terms which would (if non-zero) carry 
a periodic time-dependence, and in his paper Heisenberg says that `I could not 
prove in general that all periodic terms actually vanish, but this was the case 
for all the terms evaluated'. We do not know how many off-diagonal terms 
$W(n, n-\al)$ he did evaluate, but he clearly regarded their vanishing as a crucial test of the formalism. In Appendix B we outline the calculation of 
all off-diagonal terms for the $\lambda x^2$ force, up to order $\lambda$, as 
an example of the kind of calculation Heisenberg probably did, finishing it 
late one night on Heligoland$^{53}$. 

\section{Conclusion}

We have tried to remove some of the barriers to understanding Heisenberg's 1925 
paper, by providing (apparently for the first time) details of calculations of the type we believe he probably carried out. We hope that more people will 
thereby be encouraged to appreciate this remarkable paper. 

The fact is, Heisenberg's `amplitude calculus' works: at least for the simple 
one-dimensional problems on which he tried it out, it is an eminently 
practical procedure, requiring no sophisticated mathematical knowledge to 
implement. Since it uses the correct equations of motion, and incorporates 
the fundamental commutator (\ref{eq:2.11}) via the `quantum condition' (\ref{eq:2.16}), 
the answers obtained are completely correct, in the sense of agreeing with conventional quantum mechanics. 

We believe that Heisenberg's approach, as applied to these simple dynamical 
systems, has considerable pedagogical value, and could usefully be included 
in undergraduate courses on quantum mechanics. The multiplication law (\ref{eq:2.10}) has a convincing physical rationale, even for those who (like 
Heisenberg) do not recognize it as matrix muliplication; indeed, this piece of  quantum physics  could  provide an exciting application for those learning about matrices in a concurrent  mathematics course. The simple examples  of (\ref{eq:2.10}), in formulae such as (\ref{eq:2.22}) or the analogous one for the 
$\lambda {\hat{x}}^3$ force, introduce the student directly to the fundamental 
quantum idea that a transition from one state to the other {\em occurs via all 
possible intermediate states}, something which can take time to emerge in the 
traditional wave-mechanical approach. The solution of the quantum SHO, sketched at the end of the previous section, is simple and elementary, in comparison 
with the standard methods. Finally, the type of perturbation theory employed here provides an instructive introduction to the 
technique, being more easily related to the classical analysis than is 
conventional quantum-mechanical perturbation theory (which students tend to 
find very formal). 

It is of course true that many important problems in quantum mechanics are much 
more conveniently handled in the wave-mechanical formalism: unbound problems 
are an obvious example, but even the Coulomb problem required a famous 
{\em tour de force} from Pauli$^{54}$. Nevertheless, a useful seed may be 
sown, so that when students meet  problems involving a finite 
number of discrete states - for example, in the treatment of spin - the 
introduction of matrices will come as less of a shock. And they may enjoy 
the realisation that the somewhat mysteriously named `matrix elements' of wave-mechanics are indeed 
the elements of Heisenberg's matrices.     
 \vspace{.5in}

{\LARGE{\bf{Appendix A: The quantum condition (\ref{eq:2.16}) and $\hat{x}\hat{p} - \hat{p}\hat{x} = {\rm i} \hbar$.}}}
\vspace{.25in}

Consider the $(n, n)$ element of $(\hat{x}\dot{\hat{x}} - \dot{\hat{x}}\hat{x})$. 
This is 
\bea
&&\sum_\al X(n, n-\al){\rm i}\omega(n-\al, n)X(n-\al, n) \nonumber \\
 &&- \sum_\al {\rm i} \omega(n, n-\al) X(n, n-\al) X(n-\al, n). \label{eq:A.1}
\eea
In the first term of (\ref{eq:A.1}), the sum over $\al > 0$ may be re-written 
as 
\be
-{\rm i}\sum_{\al > 0} \omega(n, n-\al) |X(n, n-\al)|^2 \label{eq:A.2} 
\ee
using $\omega(n, n-\al) = -\omega(n-\al, n)$ from (\ref{eq:2.1}) and 
$X(n-\al, n) = X^{*}(n, n-\al)$ from (\ref{eq:3.2}), 
while the sum over $\al < 0$ becomes, similarly,  
\be 
{\rm i} \sum_{\al > 0} \omega(n+\al, n) |X(n+\al, n)|^2 \label{eq:A.3}
\ee
on changing $\al$ to $-\al$. Similar steps in the second term of (\ref{eq:A.1}) 
lead to the result 
\bea 
(\hat{x}\dot{\hat{x}} - \dot{\hat{x}} \hat{x})(n, n) &= & 2{\rm i} \sum_{\al > 0} [\omega(n+\al, n)|X(n+\al, n)|^2 \nonumber \\
&& - \omega(n, n-\al)|X(n, n-\al)|^2] \nonumber \\
&=& 2 {\rm i} h /(4 \pi m), \label{eq:A.4}
\eea
where the last step follows from the `quantum condition' (\ref{eq:2.16}). Setting $\hat{p} = m \dot{\hat{x}}$ we find 
\be 
(\hat{x}\hat{p} - \hat{p} \hat{x}) (n, n) = {\rm i} \hbar \label{eq:A.5} 
\ee
for all values of $n$. This is the result which Born found$^{36}$ shortly after 
reading Heisenberg's paper. In the further development of the theory the 
value of the `fundamental commutator' $\hat{x}\hat{p} - \hat{p} \hat{x}$, 
namely ${\rm i}\hbar$ times the unit matrix, was taken to be a basic postulate. 
The sum rule (\ref{eq:2.16}) is then derived by taking the $(n, n)$ matrix 
element of the relation $[\hat{x}, [\hat{H}, \hat{x}]] = \hbar^2/m$.   

\vspace{.5in}
{\LARGE{\bf{Appendix B: Calculation of the off-diagonal matrix elements of 
the energy $W(n, n-\al)$ for 
the $\lambda x^2$ force, up to order $\lambda$.}}}
\vspace{.25in}

We shall show that, for $\al \neq 0$,  all the elements $(n, n-\al)$ of the energy operator 
$\frac{1}{2}m {\dot{\hat{x}}}^2 + \frac{1}{2}m \omega^2_0 {\hat{x}}^2 + \frac{1}{3}\lambda m {\hat{x}}^3$ vanish,  up to order $\lambda$. 

We begin by noting the qualitative point that, at any given order in $\lambda$, 
only a limited number of elements $W(n, n-1), W(n, n-2),\dots$ will contribute, 
since the amplitudes $X(n, n-\al)$ are suppressed by increasing powers of 
$\lambda$ as $\al$ increases. In fact, for $\al \geq 2$ the leading power of 
$\lambda$ in $W(n, n-\al)$ is $\lambda^{\al -2}$, arising from terms such as 
$X(n, n-1)X(n-1, n-\al)$ and $\lambda X(n, n-1)X(n-1, n-2)X(n-2, n-\al)$. Thus 
to order $\lambda$ we need only calculate $W(n, n-1), W(n, n-2), W(n, n-3)$. 

(a) \underline{$W(n, n-1)$}

There are four $O(\lambda)$ contributions to the $(n, n-1)$ element of 
$\frac{1}{2}m \omega^2_0 {\hat{x}}^2$: 
\bea
&&\frac{1}{4}m \omega^2_0 \lambda\{a^{(0)}(n, n)a^{(0)}(n, n-1) + a^{(0)}(n, n-1)a^{(0)}(n-1, n-1) + \nonumber \\
&&\frac{1}{2}[a^{(0)}(n, n+1)a^{(0)}(n+1, n-1) + a^{(0)}(n, n-2)a^{(0)}(n-2, n-1)]\} \nonumber \\
&&=- \frac{5}{24} m \lambda \beta^3 n \sqrt{n}. \label{eq:B.1}
\eea
There are two $O(\lambda)$ contributions to the $(n, n-1)$ element of 
$\frac{1}{2}m{\dot{\hat{x}}}^2$:
\bea
&&- \frac{1}{8} \lambda m \{ \omega^{(0)}(n, n+1) \omega^{(0)}(n+1, n-1) a^{(0)}(n, n+1) a^{(0)}(n+1, n-1) + \nonumber \\
&&\omega^{(0)}(n, n-2)\omega^{(0)}(n-2, n-1)a^{(0)}(n, n-2)a^{(0)}(n-2, n-1)\} 
\nonumber \\
&&= \frac{1}{12}m \lambda \beta^3 n \sqrt{n}.
\label{eq:B.2}
\eea
There are three $O(\lambda)$ contributions to the $(n, n-1)$ element of 
$\frac{1}{3}m \lambda {\hat{x}}^3$:
\bea 
&&\frac{1}{24}m \lambda \{ a^{(0)}(n, n-1)a^{(0)}(n-1, n) a^{(0)}(n, n-1) + \nonumber \\
&&a^{(0)}(n, n-1)a^{(0)}(n-1, n-2)a^{(0)}(n-2, n-1) \nonumber \\
&& + a^{(0)}(n, n+1)a^{(0)}(n+1, n)a^{(0)}(n, n-1)\} \nonumber \\
&&= \frac{1}{8} m \lambda \beta^3 n \sqrt{n}.  
\label{eq:B.3}
\eea
The sum of (\ref{eq:B.1}), (\ref{eq:B.2}) and (\ref{eq:B.3}) vanishes, as required.

(b) \underline{$W(n, n-2)$}

The leading contribution is independent of $\lambda$. From the term $\frac{1}{2}m\omega^2_0 {\hat{x}}^2$ it is 
\be
\frac{1}{8} m \omega^2_0 a^{(0)}(n, n-1)a^{(0)}(n-1, n-2), \label{eq:B.4}
\ee
which is cancelled by the corresponding term from $\frac{1}{2}m{\dot{\hat{x}}}^2$. The next terms are $O(\lambda^2)$, for example from the leading term in the $(n, n-2)$ element of $\frac{1}{3}\lambda m {\hat{x}}^3$. 

(c)\underline{$W(n, n-3)$}

There are two $O(\lambda)$ contributions from $\frac{1}{2}m \omega^2_0{\hat{x}}^2$:
\bea
&&\frac{1}{8}m\omega^2_0 \lambda \{ a^{(0)}(n, n-1)a^{(0)}(n-1, n-3) + a^{(0)}(n, n-2) a^{(0)}(n-2, n-3)\} \nonumber \\
&&= \frac{1}{24}m \lambda \beta^3 \sqrt{n(n-1)(n-2)}.
\label{eq:B.5}
\eea
There are two $O(\lambda)$ contributions from $\frac{1}{2}m {\dot{\hat{x}}}^2$:
\bea
&&-\frac{1}{8}m \lambda \{ \omega^{(0)}(n, n-1)a^{(0)}(n, n-1)\omega^{(0)}(n-1, n-3)a^{(0)}(n-1, n-3) \nonumber \\
&& + \omega^{(0)}(n, n-2)a^{(0)}(n, n-2)\omega^{(0)}(n-2, n-3)a^{(0)}(n-2, n-3)\} \\
&& = - \frac{1}{12} \lambda m \beta^3 \sqrt{n(n-1)(n-2)}. \label{eq:B.6}
\eea
There is only one $O(\lambda)$ contribution from $\frac{1}{3}m \lambda {\hat{x}}^3$:
\bea
&&\frac{1}{24} m \lambda a^{(0)}(n, n-1) a^{(0)}(n-1, n-2) a^{(0)}(n-2, n-3) \nonumber \\
&& = \frac{1}{24} \lambda  m \beta^3 \sqrt{n(n-1)(n-2)}. \label{eq:B.7}
\eea
The sum of (\ref{eq:B.5}), (\ref{eq:B.6}) and (\ref{eq:B.7}) vanishes, as required.

\vspace{.5in}
\mbox{}$^{1}$ W. Heisenberg, ``\"{U}ber quantentheoretische Umdeutung kinematischer und mechanischer Beziehungen'', Z. Phys. {\bf 33}, 879-893 (1925).\\
\mbox{}$^2$ This is the title of the English translation of the paper which is 
included as paper 12 in Ref. 3, pp. 261-276. In the present paper we shall 
refer exclusively to this translation; in particular, we shall refer to the 
equations in it as `H(1), H(2) . . . ' etc.\\
\mbox{}$^{3}$ {\em Sources of Quantum Mechanics}, edited by B. L. van der 
Waerden (Amsterdam, North-Holland, 1967). \\
\mbox{}$^4$ M. Born and P Jordan, ``Zur Quantenmechanik'', Z. Phys. {\bf 34}, 858-888 (1925), 
reprinted in translation as paper 13 in Ref. 3.   \\
\mbox{}$^5$ P. A. M. Dirac, ``The Fundamental Equations of Quantum Mechanics'', Proc. Roy. Soc. A {\bf 109}, 642-653 (1926), 
paper 14 in Ref. 3. \\
\mbox{}$^6$ M. Born, W. Heisenberg and P. Jordan, ``Zur Quantenmechanik II'', Z. Phys. {\bf 35}, 557-615 (1926), reprinted in translation as paper 15 in Ref. 3.  \\
\mbox{}$^7$ S. Weinberg, {\em Dreams of a Final Theory} (New York, Pantheon, 1992). \\
\mbox{}$^8$ S.-I. Tomonaga, {\em Quantum Mechanics Volume 1 Old Quantum Theory} (Amsterdam, North-Holland, 1962). \\
\mbox{}$^9$ M. Jammer, {\em The Conceptual Development of Quantum Mechanics} 
(New York, McGraw-Hill, 1966). \\
\mbox{}$^{10}$ E. MacKinnon, ``Heisenberg, Models and the Rise of Matrix Mechanics'', Hist. Stud. Phys. Sci. {\bf 8}, 137-188 (1977).\\ 
\mbox{}$^{11}$ J. Mehra and H. Rechenberg, {\em The Historical Development of 
Quantum Theory} vol 2 (New York, Springer-Verlag, 1982). \\
\mbox{}$^{12}$ J. Hendry, {\em The Creation of Quantum Mechanics and the 
Bohr-Pauli Dialogue} (Dordrecht, D. Reidel, 1984). \\ 
\mbox{}$^{13}$ T.-Y. Wu, {\em Quantum Mechanics} (Singapore, World Scientific, 
1986). \\
\mbox{}$^{14}$ M. Taketani and M. Nagasaki, {\em The Formation and Logic of 
Quantum Mechanics} vol 3 (Singapore, World Scientific, 2002). \\
\mbox{}$^{15}$ G. Birtwistle, {\em The New Quantum Mechanics} (Cambridge, Cambridge University Press, 1928). \\
\mbox{}$^{16}$ M. Born, {\em Atomic Physics} (New York, Dover, 1989). \\
\mbox{}$^{17}$ J. Lacki, ``Observability, Anschaulichkeit and Abstraction: A 
Journey into Werner Heisenberg's Science and Philosophy'', Fortschr. Phys. {\bf 50}, 440-458 (2002). \\
\mbox{}$^{18}$ J. Mehra, {\em The Golden Age of Theoretical Physics} vol 2 
(Singapore, World Scientific, 2001). \\   
 \mbox{}$^{19}$ Ref. 7, page 53. \\
\mbox{}$^{20}$ All quotations are from the English translation in Ref. 3. \\
\mbox{}$^{21}$ We use $\omega$ here rather than Heisenberg's $\nu$. \\
\mbox{}$^{22}$ The reader may find it helpful at this point to consult 
Ref. 23, which provides a clear account of the connection between the `classical' analysis of an electron's periodic motion and simple `quantum' 
versions. See also J. D. Jackson, {\em Classical Electrodynamics} 2nd edtn (New York, Wiley, 1975), section 9.2. \\
\mbox{}$^{23}$ W. A. Fedak and J. J. Prentis, ``Quantum jumps and classical 
harmonics'', Am. J. Phys. {\bf 70}, 332-344 (2002). \\
\mbox{}$^{24}$ We depart from the Gothic notation of Ref. 3 (and Ref. 1), 
preferring that of Tomonaga, Ref. 8, pp 204-224. The association $X_{\alpha}(n)  \leftrightarrow X(n, n-\alpha)$ is generally true only for non-negative $\alpha$. For negative values of $\alpha$, a general term in the classical Fourier series is $X_{-|\alpha|}(n) {\rm exp}[- {\rm i} \omega(n) |\alpha| t]$. 
Replacing $-\omega(n) |\alpha|$ by $-\omega(n, n-|\al|)$, which is equal to $
\omega(n-|\al|, n)$  using  
(\ref{eq:2.1}), we see that $X_{-|\alpha|}(n) 
\leftrightarrow X(n-|\alpha|, n)$. The association $X_{-|\alpha|} \leftrightarrow X(n, n+|\alpha|)$ would {\em not} be correct since $\omega(n, n+|\alpha|)$ is not the same, in general, as $\omega(n-|\alpha|, n).$
  \\
\mbox{}$^{25}$ Conventional notation - subsequent to Ref. 4 - would replace 
`$n-\al$' by a second index `$m$', say. We prefer to remain as close as 
possible to the notation of Heisenberg's paper, for ease of reference. \\
\mbox{}$^{26}$ This step apparently did not occur to him immediately - see 
Ref. 11, page 231. \\
\mbox{}$^{27}$ Actually not quite: we have taken the liberty of changing 
the order of the arguments in the first terms in the braces; this (correct) 
order is as given in the equation Heisenberg writes before H(20). \\
\mbox{}$^{28}$ W. Thomas, ``\"{U}ber die Zahl der Dispersionelektronen, die 
einem station\"{a}ren Zustande zugeordnet sind (Vorl\"{a}ufige Mitteilung)'', 
Naturwiss. {\bf 13}, 627 (1925). \\
\mbox{}$^{29}$ W. Kuhn, ``\"{U}ber die Gesamtst\"{a}rke der von einem Zustande augehenden Absorptionslinien'', Z. Phys. {\bf 33} 408-412(1925), reprinted in 
translation as paper 11 in ref. 3. \\
\mbox{}$^{30}$ W. Heisenberg, as discussed in Ref. 11 pages 243 ff. \\
\mbox{}$^{31}$ Ref. 9, p 193; $\Phi$ is any function defined for stationary states.  \\
\mbox{}$^{32}$ M. Born, ``\"{U}ber Quantenmechanik'', Z. Phys. {\bf 26}, 379-395 (1924), reprinted in translation as paper 7 in ref. 3. \\
\mbox{}$^{33}$ Ref. 9, p 202. \\
\mbox{}$^{34}$ H. A. Kramers and W. Heisenberg, ``\"{U}ber die Streuung von Strahlen durch Atome'', Z. Phys. {\bf 31}, 681-707 (1925), reprinted in translation as paper 10 in Ref. 3. \\
\mbox{}$^{35}$ For considerable further detail on dispersion theory, sum rules, 
and the `discretisation' rules see Ref. 8 pages 142-147 and pages 206-208, and 
Ref. 9 section 4.3. \\
\mbox{}$^{36}$ See Ref. 3 page 37. \\
\mbox{}$^{37}$ For an interesting discussion of the possible reasons why 
he chose to try out his scheme on the anharmonic oscillator, see Ref. 11 pages 
232-235; and also Ref. 3 page 22. Curiously, most of the commentators - with 
the notable exception of Tomonaga (Ref. 8) - seem to lose interest in the details of the calculations at this point. \\
\mbox{}$^{38}$ The $(n, n-\al)$ matrix element, in the standard terminolgy. \\
\mbox{}$^{39}$ Equation ({\ref{eq:2.22}) 
is not in Heisenberg's paper, though it is 
given by Tomonaga, Ref. 8, equation (32.20$^{'}$). \\
\mbox{}$^{40}$ Note that this means that, in $x(t)$, `all the terms which are 
of order $\lambda^p$' arise from many different terms in (\ref{eq:2.24}). \\
\mbox{}$^{41}$ Except that Heisenberg re-labels most of the $a_{\al}$'s as 
$a_{\al}(n)$. \\
\mbox{}$^{42}$ Actually he writes $a_0(n)$ in place of $a(n, n)$ in (\ref{eq:2.40}), through an oversight. \\
\mbox{}$^{43}$ MacKinnon (Ref. 10) suggests how, in terms of concepts 
from the `virtual oscillator' model, equations (\ref{eq:2.32}) - 
(\ref{eq:2.35}) may be `transformed' into equations (\ref{eq:2.40}) - (\ref{eq:2.43}); we do not agree with MacKinnon (Ref. 10, footnote 62) 
regarding `mistakes' in (\ref{eq:2.42}) and (\ref{eq:2.43}). \\   
\mbox{}$^{44}$ In the version of the quantum condition which Heisenberg 
gives just before H(20), he unfortunately uses the same symbol for the 
transition amplitudes as in H(16) - see our (\ref{eq:2.16}) - but replaces 
`$4 \pi m$' by `$\pi m$', not explaining where the factor 1/4 has come 
from (see (\ref{eq:3.6})); he also omits the $\lambda$'s.  \\
\mbox{}$^{45}$ Heisenberg omits the superscripts. \\
\mbox{}$^{46}$ See for example L. I. Schiff, {\em Quantum Mechanics}, 3rd edtn 
New York, McGraw-Hill, 1968). \\
\mbox{}$^{47}$ This is the justification suggested by Born and Jordan in Ref. 4. \\
\mbox{}$^{48}$ It is essentially that given by L. D. Landau and E. M. Lifshitz, {\em Course of Theoretical Physics Vol. 3  Quantum Mechanics} 3rd edtn (Oxford, Pergamon, 1977), pages 67-68. \\  
\mbox{}$^{49}$ L. D. Landau and E. M. Lifshitz, {\em Course of Theoretical Physics Vol. 1 Mechanics} 3rd edtn (Oxford, Pergamon, 1976). \\
\mbox{}$^{50}$ This equation corresponds to equation (88) of Born and Jordan's 
paper, Ref. 4, in which there appears to be a misprint of 17/30 for 11/30. \\
\mbox{}$^{51}$ See Ref. 4. \\
\mbox{}$^{52}$ L. D. Landau and E. M. Lifshitz, as cited in Ref. 48, page 136. \\
\mbox{}$^{53}$ See W. Heisenberg, {\em Physics and Beyond} (London, Allen \& Unwin, 1971), page 61. \\
\mbox{}$^{54}$ W. Pauli, ``\"{U}ber das Wasserstoffspektrum vom Standpunkt 
der neuen Quantenmechanik'', Z. Phys. {\bf 36}, 336-363 (1926), reprinted in 
translation as paper 16 in Ref. 3.\\

\end{document}